\documentclass[useAMS,usenatbib]{mn2e}
\usepackage{graphicx,color,epsfig}

\newcommand{\be}{\begin{equation}}
\newcommand{\ee}{\end{equation}}

\newcommand{\apj}{ApJ}

\newcommand{\mnras}{MNRAS}
\newcommand{\aap}{A\&A}
\newcommand{\araa}{ARA\&A}
\newcommand{\apjl}{ApJL}

\newcommand{\nat}{Nature}

\newcommand{\icarus}{ICARUS}

\def\ltsima{$\; \buildrel < \over \sim \;$}
\def\simlt{\lower.5ex\hbox{\ltsima}}
\def\gtsima{$\; \buildrel > \over \sim \;$}
\def\simgt{\lower.5ex\hbox{\gtsima}}

\newcommand\mj{{\,{\rm M}_{\rm J}}}

\title[Metal loading of planets]{Metal loading of giant gas planets}

\author[S. Nayakshin]{Sergei Nayakshin\\ Department of Physics \& Astronomy,
  University of Leicester, Leicester, LE1 7RH, UK\\ {E-mail:~} {\rm
    Sergei.Nayakshin@le.ac.uk}}

\begin{document}

\date{Received}

\pagerange{\pageref{firstpage}--\pageref{lastpage}} \pubyear{2008}

\maketitle

\label{firstpage}

\begin{abstract}
One of many challenges in forming giant gas planets via Gravitational disc
Instability model (GI) is an inefficient radiative cooling of the pre-collapse
fragments.  Since fragment contraction times are as long at $10^5 -10^7$
years, the fragments may be tidally destroyed sooner than they contract into
gas giant planets. Here we explore the role of "pebble accretion" onto the
pre-collapse giant planets and find an unexpected result. Despite larger dust
opacity at higher metallicities, addition of metals actually accelerates --
rather than slows down -- collapse of high opacity, relatively low mass giant
gas planets ($M_p \simlt$ a few Jupiter masses).  A simple analytical theory
that explains this result exactly in idealised simplified cases is
presented. The theory shows that planets with the central temperature in the
range of $1000 \simlt T_{\rm c} \simlt 2000$~K are especially sensitive to
pebble accretion: addition of just $\sim 5$ to 10\% of metals by weight is
sufficient to cause their collapse. These results show that dust grain physics
and dynamics is essential for an accurate modelling of self-gravitating disc
fragments and their near environments in the outer massive and cold
protoplanetary discs.
\end{abstract}


\section{Introduction and background}\label{sec:intro}

Gravitational disc Instability (GI) theory for giant planet formation
\citep[e.g.,][]{Kuiper51,CameronEtal82,Boss97,Boss98} posits that
gravitational instability of the disc leads to formation of self-gravitating
gas fragments that later contract into present day planets. This view has been
strongly challenged in the last decade since it was shown that protoplanetary
discs do not cool rapidly enough to fragment onto gas clumps inside several
tens to a hundred AU
\citep{Gammie01,MayerEtal04,Rice05,Rafikov05,DurisenEtal07,SW08,Meru10}. This
would preclude the model from explaining most of the giant planets, since most
are detected at separations much smaller than this, all the way down to $\sim
0.05$~AU \citep[e.g.,][]{MQ95}.

However, \cite{Nayakshin10c} argued that giant planets could be born far out
but then migrate inward arbitrarily close to the parent star due to
gravitational torques of the disc
\citep[e.g.,][]{LinPap79,GoldreichTremaine80}. Crucially, GI planets are born
as fluffy (gas density $\rho \sim 10^{-13}$~g/cm$^3$) and cold (T$\sim 100$~K)
molecular gas fragments. In order to become dense Jupiter-like giant planets,
they must first contract to $T\simgt 2500$~K at which point a rapid dynamical
collapse occurs \citep[e.g.,][]{Bodenheimer74}. Tantalisingly, if the planets
migrate inward more rapidly than they contract, then they may be tidally
disrupted. The disruption leaves behind a rocky/icy core if grains inside the
planet managed to grow and sediment down into a core rapidly enough
\citep{McCreaWilliams65,BoleyEtal10}.  This scheme, named "Tidal Downsizing"
by \cite{Nayakshin10c}, may {\it potentially} explain any mass planets at
arbitrary separation from the host star within a single framework
\citep{ForganRice13b}.

Here we focus on giant gas planets within the GI/TD framework
specifically. While the rapid \citep[migration time $t_{\rm mig} \sim 10^4$
  years, see, e.g.,][]{BaruteauEtal11,Nayakshin10c} inward migration and tidal
disruption process of gas fragments is crucial to TD as a way of accounting
for terrestrial-like planets, this processes challenges the formation of the
gas giants themselves \citep{ZhuEtal12a}.  Firstly, since radiative
contraction time of even the Solar metallicity clouds is quite long
\citep[$\sim 10^6$ years, see, e.g.,][]{BodenheimerEtal80,VazanHelled12}, the
disruption process appears to be too efficient (see \S \ref{sec:under}), so it
is hard to explain how {\em any} close-in giant planets survive. Secondly,
radiative contraction of planets is slowed down at higher grain opacities
\citep{HB11}. Thus, tidal disruption in metal-rich environments should be even
more efficient. This would seem to contradict the well known fact that giant
planets are more abundant at higher metallicities
\citep[e.g.,][]{FischerValenti05,WangFischer14}. If this is true then close-in
giant planets must form via Core Accretion rather than GI \citep[see,
  e.g.,][]{Boley09}.

{There may appear to exist a simple solution to the conundrum of too long a
  radiative cooling time for GI clumps: a strongly decreased grain opacity due
  to grain growth. \cite{HB11} find that in models that include grain growth,
  opacity decrease cuts the planet contraction time scale to as short as $\sim
  10^3$ years. We do not favour this solution for two reasons.  Firstly, this
  would predict that protoplanetary discs of low metallicity stars would be
  ideal sites for gas giant planet formation. Instead, \cite{FischerValenti05}
  shows that low metallicity stars are the least likely to host a giant
  planet. Secondly, models of \cite{HB11} do not include collisional
  fragmentation of dust grain aggregates which are found to occur at
  relatively modest velocities in laboratory experiments
  \citep[e.g.,][]{BlumWurm08}. \cite{DD05} consider grain growth and opacity
  in protoplanetary discs of T Tauri stars. They find that models that neglect
  grain growth deplete the small grain population by a factor of a million in
  a few thousand years (see their figure 3), reducing the dust opacity
  significantly. However, the spectra of T Tauri sources clearly require a
  copious presence of small dust grains even at ages of a few Million
  years. To explain this, \cite{DD05} argue that high speed dust aggregate
  fragmenting collision must be also taken into account. This then results in
  a quasi steady-state for dust distribution in which the smallest grain
  abundance is suppressed only mildly even after about a Million years of
  grain growth (see their Fig. 7). For these two (observationally backed up)
  reasons, we discount the possibility that grain opacity in GI fragments is
  much lower than the interstellar one, although we do explore some grain
  opacity reduction below.

The solution to the too-slow-cooling conundrum that is proposed here is based
on another unexpected effect that grains have on GI clumps.}
\cite{NayakshinEtal14a} investigated the structure of the gas ``atmosphere''
around the massive solid core built within the TD gas fragments by grain
settling. Similar to CA theory, it was found that there exists a critical core
mass, $M_{\rm crit}$, beyond which the atmosphere may become self-gravitating
and too massive to be in hydrostatic balance
\citep{Mizuno80,Stevenson82}. Following CA arguments, the authors argued that
the atmosphere collapse sets off a phase of a rapid gas accretion onto the
dense core.  This may initiate a hydrodynamical collapse (termed Core-Assisted
Gas Capture instability; CAGS hereafter) of the whole gas fragment {\it
  before} a core-less fragment of same configuration would have collapsed,
thus providing a second and an unexplored way for the formation of giant
planets in the TD picture.

The original goal of this project was to follow up the work of
\cite{NayakshinEtal14a} with a hydrodynamical rather than hydrostatic code,
and to model the whole gas clump rather than only its centre. In addition, we
included external large grain deposition onto the planet, the so-called
"pebble accretion" \citep[e.g.,][]{JohansenLacerda10,OrmelKlahr10} of grains
on massive bodies embedded in protoplanetary discs. In doing so we discovered
that planets accreting grains can collapse more rapidly than those of fixed
metallicity. It turned out that this effect has nothing to do with formation
of the core directly, and instead originates in the fact that grain accretion
on a planet in itself is a form of cooling. Pending detailed analysis below,
this last statement can be understood qualitatively as follows.

Consider the total energy of the pre-collapse planet of mass $M_{\rm p}$ and
radius $R_{\rm p}$, $E_{\rm tot} \sim - GM_{\rm p}^2/R_{\rm p}$. It evolves in
time according to
\begin{equation}
{d E_{\rm tot}\over dt} = - L_{\rm rad} - {G M_{\rm p} \dot M_z\over R_{\rm p}}\;,
\label{equation1}
\end{equation}
where $L_{\rm rad}$ is the radiative luminosity of the planet, and the last
term on the right is the change in the gravitational potential energy of the
planet, $E_{\rm grav}$, due to grain accretion on it at the rate $\dot
M_z$. In the constant metallicity case (no pebble accretion) that was so far
studied in literature \citep{HB11}, the second term on the right is absent, so
that radiation is the only way for the planet to loose excess energy and
contract towards H$_2$ collapse. Increasing metallicity of the planet
increases dust opacity and decreases $L_{\rm rad}$, hence slowing down its
contraction. However, if $\dot M_z > 0$, then the planet "cools" due to
addition of metals, since $E_{\rm tot}$ becomes more negative. Deposition of
pebbles into a planet from outside may hence speed up rather than delay giant
planet formation.

The goal of this paper is to study the contraction of pre-collapse giant
planets due to grain accretion in detail. To not over-complicate this first
study by having to also study GAGC, we turn off grain growth in the planet,
with the result that grain sedimentation is negligible, and no core forms in
the planet's centre. The more general case is to be studied in a near future
paper.  We use a 1D spherically symmetric radiation hydrodynamics code as well
as analytic arguments to understand the planet's response to pebble deposition.

The paper is structured as follows. In \S \ref{sec:prelim}, the cooling
challenge of GI-born fragments is described, and the rate of pebble accretion
from the surrounding protoplanetary disc is estimated. In \S \ref{sec:methods}
and in Appendix \ref{appendix1}, numerical methods employed here are
presented.  \S \ref{sec:intro_runs} compares the evolution of planets enriched
by metals either at birth or by pebble accretion, and \S \ref{sec:theory}
outlines an analytical toy model that helps to understand why metal accretion
accelerates planet contraction. The analytical model is compared with
grain-dominated contraction of numerically integrated polytropic clouds with
fixed specific heats ratios ($\gamma$), as well as a more realistic molecular
planet in \S \ref{sec:real}. The conditions under which pebble accretion
dominates planet contraction are delineated in \S \ref{sec:grain_dom}. A
discussion is given in \S \ref{sec:discussion}.

\section{Preliminaries}\label{sec:prelim}

\subsection{The cooling challenge}\label{sec:under}

The properties of the first gas condensations in a self-gravitating disc
should be similar \citep{Nayakshin10a} to that of the ``first cores''
\citep{Larson69} in star formation, see also
\cite{Bodenheimer74,BodenheimerEtal80}.  The central temperature of these may
be as low as $\simlt 100$~K, Hydrogen is molecular, and the mass of the
fragment is of the order of a Jupiter mass to perhaps $10 \mj$
\citep{BoleyEtal10,ForganRice11,TsukamotoEtal14}.  If the fragments cool
sufficiently rapidly, reaching the central temperature of $\sim 2500$~K {\em
  before} they are tidally disrupted, then dissociation of molecular hydrogen
occurs.  Since post-collapse fragments are $\sim 3-5$ orders of magnitude
denser than pre-collapse ones, post-collapse fragments are much more likely to
survive as giant planets. H$_2$ collapse is hence a necessary step to
formation of a gas giant planet.

\begin{figure}
\psfig{file=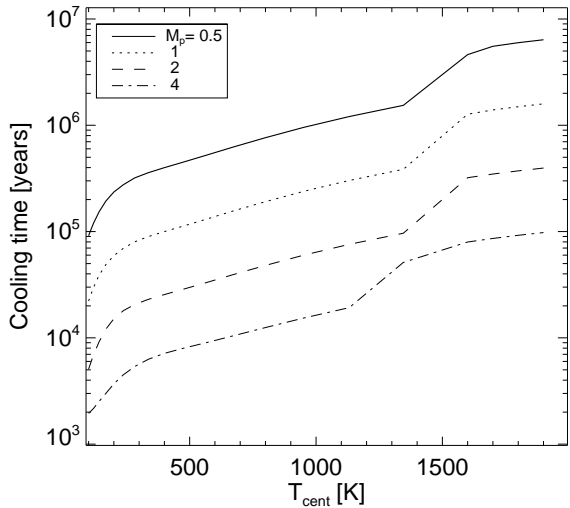,width=0.5\textwidth,angle=0}
  \caption{Radiative cooling times of pre-collapse planets of four different
    masses composed of a Solar mix of H/He and metals, as labelled on the
    figure, versus the central temperature of the planet, $T_{\rm c}$. Note
    that low mass planets are the least capable of contracting rapidly, and
    that the slowest evolution always occurs at $T_{\rm c}\simgt 1000$~K.
    Interstellar dust opacity is assumed for this figure.}
 \label{fig:tcool}
\end{figure}

However, radiative cooling times of isolated Jupiter-mass pre-collapse
fragments are $t_{\rm rad}\sim 10^6$ years \citep[e.g.,][]{BodenheimerEtal80},
far longer than the disc migration times, $t_{\rm mig} \sim 10^4$ years
\citep[e.g.,][]{BaruteauEtal11,Nayakshin10c}. Figure \ref{fig:tcool} shows the
radiative cooling time of the pre-collapse planets, defined as
\begin{equation}
t_{\rm rad} = -{E_{\rm tot}\over 2 L}\;,
\label{trad}
\end{equation}
where $E_{\rm tot}$ is the total energy of the fragment, and $L$ is the
luminosity of the planet. The factor $2$ in the denominator is introduced
empirically to better match the numerical calculations of planet contraction
 (all the way to H$_2$ dissociation) reported below.

In accord with previous results \citep{BodenheimerEtal80,HB11}, figure
\ref{fig:tcool} shows that more massive planets cool more rapidly. Also note
that, for all planetary masses, the initial cooling phase at $T_{\rm c}$ of a
few hundred K is much more rapid when the one at $T_{\rm c}\simgt
1000$~K. This is because the modulus of the total energy of the planet,
$|E_{\rm tot}|$, increases, whereas the luminosity $L(T_{\rm c})$ usually
drops with increasing $T_{\rm c}$ since the planet contracts and opacity in
the atmosphere rises. This implies that the cooling "bottle neck" for planets
of any mass is always at $T_{\rm c} \simgt 1000$~K, that is, the planets are
expected to spend more time contracting from $T_{\rm c}\sim 1000$~K to H$_2$
dissociation ($T_{\rm c}\sim 2200$~K or so, for most of our runs below) than
they spend contracting from their initial state towards $T_{\rm c}\sim
1000$~K.

Based on figure \ref{fig:tcool}, the expected outcome of GI fragment
contraction, while it also migrates inward rapidly, is that the fragment is
tidally disrupted before it collapses \citep{BoleyEtal10,Nayakshin10c}.  For
example, for $M_{\rm p} = 2\mj$, it takes more than $10^5$ years for this
collapse to take place. Furthermore, calculations by
\cite{CameronEtal82,VazanHelled12} show that irradiation from the parent star
and the surrounding disc slows the contraction of the fragments further, and
may in fact even reverse the heat flow from the planet. In this case the
planet puffs up rather than contracts with time. In the presence of tidal
shear from the parent star, this is clearly a receipt for destruction rather
than formation of a giant planet-to-be.

\subsection{The role of metals}\label{sec:metal_role}

Giant planets are observed to be over-abundant in metals
by a factor of a few to ten in the Solar System and beyond
\citep{MillerFortney11}. \cite{HB11} have shown that contraction of
pre-collapse giant planets is slowed down further at high metallicites if dust
opacity is proportional to the metallicity of the gas. 

In the context of GI/TD models, we see two principal modes of metal enrichment
of pre-collapse fragments. Firstly, fragments can be enriched by solids at
birth by efficient aerodynamic capturing of grains into spiral arms of the
gravitationally unstable disc before it collapses onto fragments
\citep[e.g.,][]{BoleyEtal11a}. In this model, therefore, pre-collapse planets
are born more metal rich than the surrounding disc. For simplicity, we then
assume that the metallicity of such planets remain constant with
time. Secondly, metals can be gained by accretion from the disc, as we explain
now.

\subsection{Accretion of metals from the disc}\label{sec:mdot_grain}

\cite{JohansenLacerda10} and \cite{OrmelKlahr10} pointed out that "pebbles",
that is small solid bodies of $\sim 10$~cm-size can accrete efficiently on
planetesimals and rocky protoplanets embedded in a disc. This can potentially
speed up assembly of massive solid cores at tens of AU distances from the host
star, where core accretion via planetesimal capture is inefficient, and solve
the problem of too long core assembly time scale of CA theory
\citep[e.g.,][]{HB14} for planets such as Neptune and Uranus.

In this picture, the accretor follows a Keplerian circular orbit of radius
$a_p$ around the star of mass $M_*$, whereas the gas in the disc moves at a
velocity slightly smaller than the local Keplerian, and there is also the
local Keplerian shear, so that there is always gas streaming past the
accretor. The gas itself does not accrete onto the accretor since the gravity
of the latter is insufficient to overcome the pressure gradient force of the
gas. The small grains, on the other hand, are not supported against accretion
onto the protoplanet by a pressure gradient.

A similar situation may hold around a much more massive molecular gas
fragment.  \cite{NayakshinCha13} show that radiative pre-heating of the
surrounding disc material by the radiation from a young embedded protoplanet
is important in deterring gas accretion onto relatively low mass
proto-planets, $M_{\rm p}\simlt 6\mj$. Such planets are found to build a
hydrostatic-like atmosphere around themselves. The gas pressure gradient
around them is large enough to prevent accretion of gas from the
disc. Fragments with initial masses $M_{\rm p}\simlt 6\mj$ are found to
migrate inward rapidly at more or less constant fragment mass until they are
tidally challenged and eventually destroyed at $R\sim 20-30$~AU. More massive
protoplanets instead accrete gas rapidly, becoming proto brown dwarfs and
stalling at about their initial locations ($R\sim 80$~AU). These results imply
that, as far as planet formation is concerned, we need to limit our attention
to relatively low mass gas fragments of a few Jupiter masses as more massive
fragments may form low mass stars, and eventually turn the system into a
stellar binary, a situation that is well outside the scope of this paper.

\cite{LambrechtsJ12} recently presented analytical estimates and numerical
simulations of pebble accretion on large planetesimals and solid cores. We
rescale their results to the case of a pre-collapse giant planet embedded in a
protoplanetary disc.  Due to the substantial mass of the planet (compared to
planetesimals or solid cores), the accretion of pebbles on it is always in the
"Hill's regime", when pebbles are accreted from the whole of the Hill's
radius, $R_{\rm H} = a_p (M_{\rm p}/3M_*)^{1/3}$, where $a_p$ is the
planet-star separation, and $M_*$ is the mass of the host star. The maximum
rate, $\dot M_{\rm z}$, at which the grain particles can be accreted by the
planet in the this regime is \citep[see eq. 38 in ][]{LambrechtsJ12}
\begin{equation}
\max \dot M_z = \dot M_{\rm H} \approx 2 R_{\rm H} \Sigma_{\rm dust} v_{\rm H}\;,
\label{mdot_hill}
\end{equation}
where $\Sigma_{\rm dust}$ is the pebbles surface density in the
protoplanetary disc, Hill's velocity is $v_{\rm H} = \Omega_p R_{\rm H}$, and
$\Omega_p$ is the Keplerian angular speed at the location of the planet. In
reality only grains intermediately strongly coupled to the gas, such that
dimensionless stopping time, $\tau_f \equiv t_f \Omega_p$, $0.1\simlt \tau_f
\simlt 1$ are accreted as efficiently as equation \ref{mdot_hill}
suggests. The physical size of the particles in the strongly coupled
  regime, that is ``pebbles'' in our definition, depends on the protostellar
  disc properties and the radial distance from the star, $R$. For the Minimum
  Mass Solar Nebula disc, \cite{LambrechtsJ12} show that particles of size $a
  \sim \hbox{a few cm}\;\times R_{1}^{-3/2}$ are in the pebble regime, where
  $R_1 = R/(10$~AU). Thus, at the outer disc regions, $R\sim 100$~AU, pebble
  regime grains are actually about 1 mm or so.

Let us define the planet metallicity doubling time scale $t_{\rm z}$ as
\begin{equation}
{1\over t_{\rm z}} \equiv {\dot M_{\rm z} \over M_{\rm z}(0)}\;,
\label{tz0}
\end{equation}
where $M_z(0) \equiv z_0 M_{\rm p}$ is the initial mass of metals in the
planet. Let the fraction of mass in the  pebble regime, that is, in the
maximum efficiency accretion regime (compared to the total mass in grains in
the disc at radius $a_p$) be $f_{\rm max}< 1$. Further, $\Sigma_{\rm dust} =
z_0 \Sigma \sim z_0 M_{\rm disc}/(\pi a_p^2)$, where $\Sigma$ is the total
(dust plus gas) surface density of the disc, $M_{\rm disc}$ is the disc mass
at radius $a_p$, and we assumed that the disc metallicity is equal to the
initial metallicity of the planet, $z_0$. The end result is,
\begin{equation}
t_{\rm z} \approx {1\over 2 f_{\rm max}} {M_{\rm p}^{1/3} M_*^{2/3} \over
  M_{\rm disc}} {2\pi \over \Omega_p}\;.
\label{tz2}
\end{equation}
Now, in the early massive protoplanetary disc stage that is of interest here, the
disc can be as massive as $0.1 M_*$ or more. Picking this value for $M_{\rm disc}$ and
$M_{\rm p} = 1 \mj$, we then have
\begin{equation}
t_{\rm z} \sim 10^3 \hbox{ years } {1\over 2 f_{\rm max}} {\left(a_p\over 100
  AU\right)}^{3/2}\;.
\label{tz3}
\end{equation}
This shows that if the population of large grains is significant, e.g.,
$f_{\rm max}\simgt 0.1$, then the time scale to double the initial metal
content of the gas fragment is comparable or shorter than the typical
migration time, $t_{\rm mig}\sim 10^{4}$~years \citep{Nayakshin10c}. Tidal
Downsizing planets may then be non-trivially over-abundant in metals by the
time they migrate into the inner few AU of the protoplanetary disc region.

Clearly, $\dot M_{\rm z}$ and $t_{\rm z}$ are functions of the disc
properties, of the planet's location within the disc. Further, the disc itself
may be influenced by the interaction with the planet. For now we explore the
case of a fixed $t_{\rm z}$, in which the planet is loaded by pebbles (metals)
from outside at a constant rate as given by equation \ref{tz0}. The value of
$t_{\rm z}$ is varied in a broad range to study the parameter space.

\section{Numerical methods}\label{sec:methods}

The code used here is a spherically symmetric Lagrangian hydrodynamics code
first described in \cite{Nayakshin10a}, expanded and updated since then as
detailed in \cite{Nayakshin10b,Nayakshin11b,Nayakshin14b}.  Here we actually
simplify the dust-gas interactions module of the code to expose the
metal-loading effect most clearly. The simplified set of equations permits a
comparison with analytical solutions, which serves as a check of both the code
and our physical intuition. In the papers cited just above, the focus was on
formation of dense metal cores inside the fragment, and hence the relative
dynamics of gas and pebbles within the fragment was important and modelled by
two-fluid equations with aerodynamical gas force coupling the two species. In
this paper, we turn off grain growth and vaporisation physics, assuming that
grains are always "small", setting $a=1 \mu$m. In practice this means that
there is no relative motion of gas and grains and hence our equations reduce
to the standard one-fluid gas-dynamical equations. Physically, this limit
corresponds to the case when grains deposited into the outer reaches of the
planet either do not grow sufficiently rapidly or get vaporised quickly if the
fragment is too hot (e.g., water ice is vaporised already at $T\sim
150$~K). The grains can however be mixed in with the gas throughout the cloud
quickly by convection and turbulence
\citep[e.g.,][]{HelledEtal08,Nayakshin11b,HB11}, so that the metallicity of
the planet (related to the grain to gas mass ratio) is set to be uniform
inside the planet at all times. The more complicated situation, where a
relative gas-grain motion does occur, is of course of a significant interest
as well, but, due to a considerably expanded parameter space in that case
(e.g., has the core formed or not, and what is the chemical composition of
that core?) it is to be presented in the near future elsewhere. We emphasise
that the ``metal loading'' effect presented here exists when the grains are
allowed to sediment as well; it is however not possible to study that
analytically since grains are redistributed within the planet in a complicated
way.

Following \cite{HB11}, we make a reasonable assumption that dust opacity in
the fragment is directly proportional to the metallicity of the gas, $z$. The
opacity $\kappa(\rho, T)$ is then given by 
\begin{equation}
\kappa(\rho, T) = f_{\rm op} \kappa_0(\rho, T) {z \over z_\odot}\;,
\label{kappa0}
\end{equation}
where $\kappa_0(\rho, T)$ are the interstellar gas plus dust opacities from
\cite{ZhuEtal09} which assume Solar metallicity, $z_\odot$, and $f_{\rm
  op}=$~const~$\le 1$ is a positive constant, set to 0.1 everywhere in this
paper, to account for possible opacity reduction due to grain growth. The main
conclusions of our paper are unchanged irrespectively of the value of $f_{\rm
  op}$, except that the parameter space where metal loading dominates as a
cooling process of course does depend on $f_{\rm op}$, as described below in
\S 7.

The full set of equations being solved for numerical experiments in this paper
is presented in the Appendix. The initial conditions are polytropic spheres of
a given central temperature $T_{\rm c}$, metallicity (usually Solar
metallicity), and total mass $M_{\rm p}$.

\section{The metal loading "paradox"}\label{sec:intro_runs}

\begin{figure*}
\begin{minipage}{3.2in}
\psfig{file=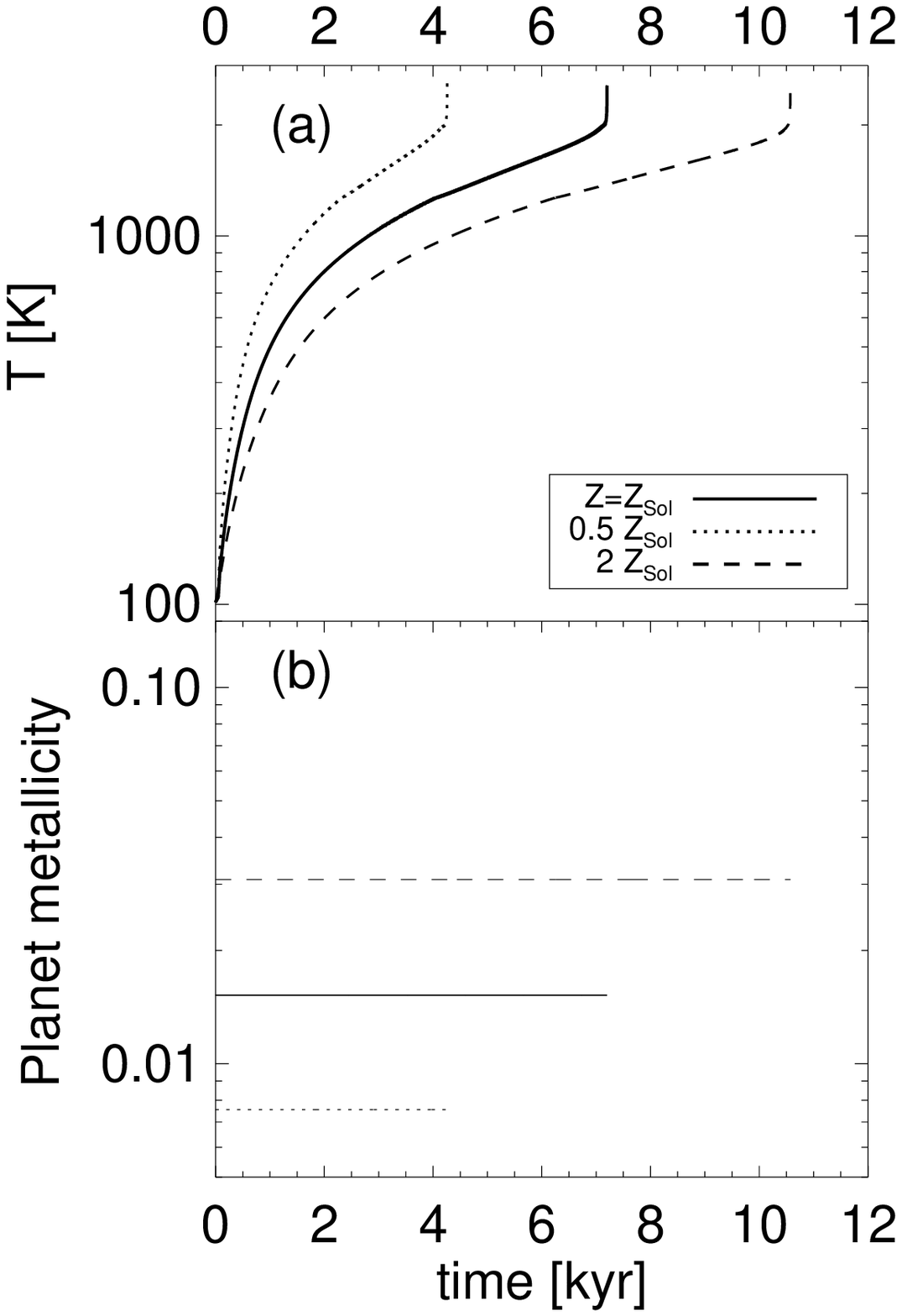,width=0.99\textwidth,angle=0}
\end{minipage}
\begin{minipage}{3.2in}
\psfig{file=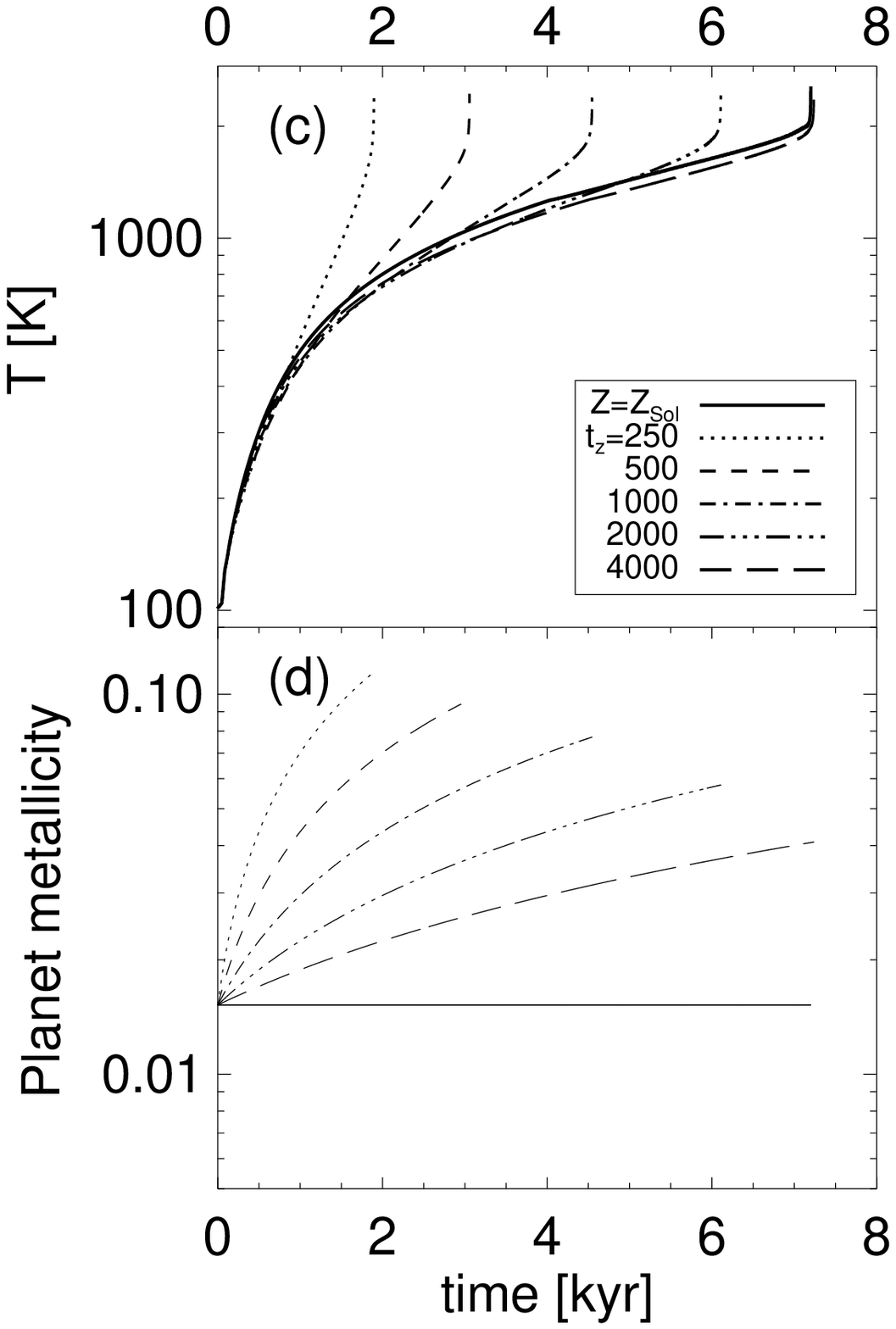,width=0.99\textwidth,angle=0}
\end{minipage}
  \caption{Top left panel (a): evolution of the central temperature versus
    time for constant metallicity planets of $4 \mj$ masses. The bottom left
    panel (b) shows the (constant in time) metallicity, $z$, of the
    planets. Right panels (c) and (d): same but for planets loaded by grains
    at constant rates parameterised by the metallicity doubling time
    $t_z$. Note that the faster the metals are added to the planet, the
    quicker it collapses.}
 \label{fig:exp1}
\end{figure*}

Figure \ref{fig:exp1} shows contraction calculations for a planet of $M_{\rm
  p} = 4 \mj$ masses, with grain opacity reduced by a factor of $10$,
performed under two different assumptions about the planet's grain
content. The left hand panels show the cases of constant planet metallicity
(enrichment at birth model), where $z = 0.5, 1$ and 2 times the Solar
metallicity.  The right hand panels show the same calculation but for the
planets all starting with the Solar grain abundance at $t=0$ and then
metal-loaded by pebble deposition at five different (constant) rates as
labelled on the figure. In particular, the curves are computed for metal
loading times ranging from $t_z = 250$~years to $t_z = 4000$~years, in steps
of a factor of 2.  Note that the shortest $t_z$ values are unlikely to be
reached even in the inner disc regions \citep[unless radial drift of pebbles,
  see][significantly increases the local pebble abundance]{LJ14}. However,
such cases expose the metal loading effects clearly. In a follow up paper
(Nayakshin 2014, submitted), the pebble accretion rate is calculated
self-consistently from the surrounding disc properties, which is modelled in
some detail similarly to \cite{NayakshinLodato12}.

The constant metallicity cases, panels (a) and (b), are in a qualitative
agreement with results of \cite{HB10}. The time taken by the planet to
contract to the point of collapse increases with increasing $z$, somewhat less
rapidly than a linear proportion.

Panels (c) and (d) show a completely different evolution, however. The more
rapidly the metals are added to the planet, the more rapidly it contracts and
eventually collapses via H$_2$ dissociation. In the most rapid metal-loading
case, $t_z = 250$~years, the collapse time scale is shortened by a factor of
almost four compared with the Solar metallicity case. For this particular
case, the metallicity at collapse is $z\approx 0.12$, eight times larger than
the starting value, $z_0 = 0.015$. Based on the left hand side panels, one
would expect that collapse time scale for such a high metallicity fragment
would be a factor of $\sim 5$ longer than the solar metallicity case, not
shorter. There is thus over an order of magnitude difference in the collapse
time scale for this particular case between the enrichment at birth and a
gradual metal deposition models.

\begin{figure}
\psfig{file=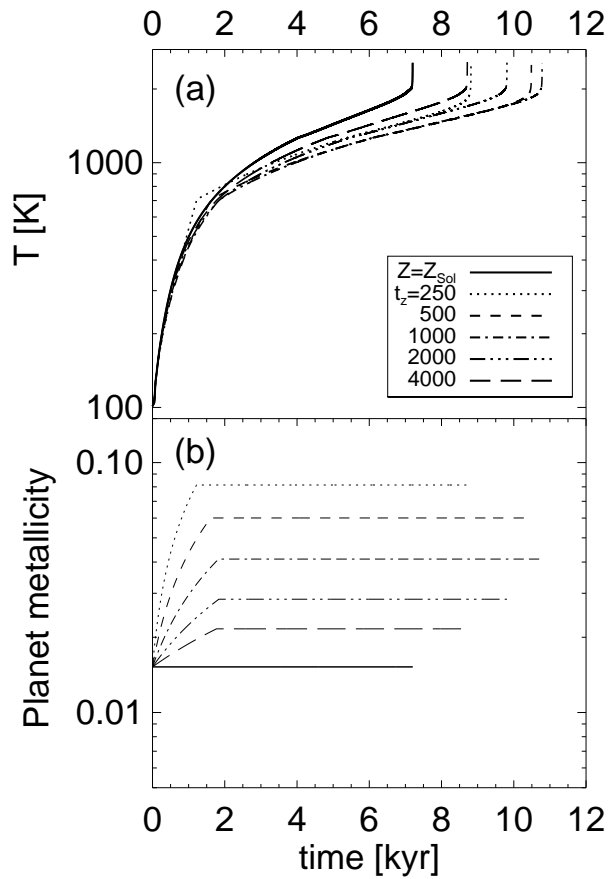,width=0.5\textwidth,angle=0}
  \caption{Same as the right hand side panels of figure \ref{fig:exp1} but
    grain accretion is (arbitrarily) turned off once the central temperature
    of 700 K is reached. Note that the faster the metals are loaded, the
    faster the fragments contract, but once grain accretion stops the
    situation reverses; the more metals are in the planet, the slower it
    contracts.}
 \label{fig:exp2}
\end{figure}

Figure \ref{fig:exp2} shows another numerical experiment that helps to shed
more light on the situation. The planets here are started exactly as the ones
presented in figures 2c and 2d, with the same choices for the metal doubling
time, but grain accretion is termined when the central temperature reaches
$T_{\rm c} =700$~K. The exact numerical value of the termination temperature
is not important. Figure \ref{fig:exp2} shows that once grain accretion stops,
contraction of the cloud continues {\it at a slower} rate than that for a
non-metal-polluted planet, i.e., one recovers the main result of the fixed
metallicity cases shown in figure \ref{fig:exp1}a and \ref{fig:exp1}b. For
example, for the case of the most rapid metal loading, $t_z = 250$~years, the
fragment collapsed at less than 2,000 years in figure \ref{fig:exp1}(a), but
collapses only at $t\approx 9,000$~years when grain accretion is discontinued.

These results indicate that pebble deposition into a pre-collapse planet
produces two opposing effects, one delaying contraction of the planet, and the
other speeding it up.  The former effect is due to the increase in the dust
opacity which slows down radiative contraction of the fragment. In our opacity
model, this effect depends on the instantaneous amount of dust in the planet,
that is, its metallicity. The latter effect however depends not on the amount
of dust in the planet but on the rate of grain deposition into the planet, as
hinted in the Introduction.

\section{A toy model}\label{sec:theory}

To understand how grain accretion can accelerate collapse of a planet we turn
to a simple analytical model in which the planet is modelled as a polytropic
sphere of adiabatic index $\gamma = 1 + 1/n$. The equation of state (EOS) of
the gas in the sphere is
\begin{equation}
P = K \rho^{\gamma}\;,
\label{eos0}
\end{equation}
where $P$ and $\rho$ and the total (gas plus metals) pressure and density,
respectively, and $K$ is a constant throughout all of the planet. Constancy of
$K$ is assumed to be maintained by convection that is known to be the main
energy transfer mechanism within H$_2$ dominated planets
\citep[e.g.,][]{HS08}.

The theory is only approximate, since there is no single value of $\gamma$
that could describe a planet dominated by molecular hydrogen exactly. For gas
dominated by molecular hydrogen, $\gamma$ varies from $\gamma = 5/3$ at
$T\simlt 100$~K to $\gamma = 7/5$ in a relatively broad temperature range,
$200 \simlt T \simlt 1000$~K, and finally drops to $\gamma \approx 1.1-1.2$ at
$T\simgt 1500$~K \citep[see figure \ref{fig:H2_planets}a below,
  and][]{BoleyEtal07}. An H$_2$ dominated gas fragment spans a range of
temperatures, from the maximum at the centre, $T_{\rm c}$, to the minimum at
the atmosphere of the planet, $T_{\rm eff}$, which may be as low as tens of K
\citep{VazanHelled12}, so $\gamma$ varies within the planet
significantly. However, qualitatively, H$_2$ dominated gas fragments behave as
polytropes with $\gamma$ varying from $5/3$ at low $T_{\rm c}$ to $\gamma< 4/3$
at $T_{\rm c}\simgt 2000$~K.

The total energy of a polytrope of mass $M_{\rm p}$ and radius $R_{\rm p}$ is
\citep[e.g.,][]{Chandrasekhar57}
\begin{equation}
E_{\rm tot} = - {3-n\over 5-n} {G M_{\rm p}^2\over R_{\rm p}} \equiv -{3\gamma - 4\over
  5\gamma- 6} {G M_{\rm p}^2\over R_{\rm p}}\;,
\label{etot1}
\end{equation}
For $\gamma<4/3$, the total energy of the planet changes sign, which marks the
dynamical instability to gravitational collapse for gas with $\gamma<4/3$. In
the context of molecular hydrogen dominated fragments, the instability leads
to H$_2$ molecules dissociation and the gravitational collapse of the planet.

Consider now an instantaneous addition of grains of mass $\delta M_z\ll M_z
\ll M_p$ throughout the planet, where $M_z$ is the current mass of metals in
the planet, so that grain mass added within radius $R$ is $\delta M =
(M/M_{\rm p}) \delta M_z $, where $M$ is planet's mass interior to $R<R_{\rm
  p}$. If grains are not vaporised, then the thermal energy of the gas/grain
mixture does not vary appreciably due to this addition (kinetic energy in
Brownian motions of grains is negligible since the weight of a grain is much
larger than that of H$_2$ molecule). However, there is a change in the
gravitational potential energy of the planet,
\begin{equation}
\delta E_{\rm grav} = -\int_0^{M_p} \left( G {M+\delta M \over R} \left[d M+d
  \delta M\right] - {GM\over R} dM\right) \;.
\label{delta_grav1}
\end{equation}
Assuming that $R$ does not vary during this instantaneous mass variation, and
since $\delta M/M = \delta M_z/M_{\rm p} =$~const,
\begin{equation}
\delta E_{\rm grav} =  2 {\delta M_z\over M_p} E_{\rm
  grav}\;,
\label{delta_grav2}
\end{equation}
where $E_{\rm grav} = - \int (GM/R) dM = -(3/(5-n)) G M_{\rm p}^2/R_{\rm
  p}$. Due to energy conservation, $\delta E_{\rm tot} = \delta E_{\rm grav}$,
and hence total energy of the planet evolves according to
\begin{equation}
- {3-n\over 5-n} {d \over dt} {G M_{\rm p}^2\over R_{\rm p}} = - {6\over 5-n}
{G M_{\rm p}^2\over R_{\rm p}} \;{\dot M_z \over M_{\rm p}} \;.
\label{etot2}
\end{equation}
Now, since grain accretion is the only mass gain term for the planet, $\dot
M_p = \dot M_z$, and so equation \ref{etot2} integrates to
\begin{equation}
\ln {G M_{\rm p}^2\over R_{\rm p}} = {6\over 3-n}\ln M_{\rm p} +
\hbox{const}\;.
\end{equation}
It is apparent that as $M_{\rm p}$ increases, $| E_{\rm tot}|$ increases as well
(and rapidly if $n\rightarrow 3$).  If $M_{\rm p}^0$and $R_{\rm p}^{0}$ are
the initial planet mass and radius before grain accretion sets in, then the
constant in the above equation can be eliminated, resulting in
\begin{equation}
{G M_{\rm p}^2\over R_{\rm p}}= \left({G M_{\rm p}^2\over R_{\rm p}}\right)_0
\left[{M_{\rm p}\over M_{\rm p}^0}\right]^{6 \over (3-n)}
\label{etot3}
\end{equation}
The planet's mass is the sum of the mass of the gas and that of metals,
$M_{\rm p} = (1-z) M_{\rm p} + z M_{\rm p}$, and $M_{\rm gas} = M_{\rm p}
(1-z)=$~const. Thus, $M_{\rm p}/M_{\rm p}^0 = (1-z_0)/(1-z)$, where $z_0$ is
the initial metallicity of the planet. Equation \ref{etot3} can be now
re-written in terms of changing metallicity of the planet rather than its
total mass.

Provided that the mean molecular weight of the gas, $\mu$, is known, one can
find the central temperature of the polytrope,
\begin{equation}
T_{\rm c} =  A_n {GM_{\rm p}\mu \over k_{\rm b} R_{\rm p}}
\label{tc0}
\end{equation}
where $A_n = - [(n+1) \xi_1\theta'_n]^{-1} \sim 1$ is a function of $n$ but
not $M_{\rm p}$ or $R_{\rm p}$. For $n=5/2$, $A_n \approx 0.7$. Since we
neglect metal's contribution to pressure and internal energy of the mix and
assume a homogeneous spreading of metals inside the planet, $\mu \approx \mu_0
M_{\rm p}/M_{\rm g} = \mu_0/(1-z)$, where $\mu_0$ is the mean molecular weight
of H$_2$/He mixture. From equation \ref{tc0}, $T_{\rm c}\propto GM_{\rm
  p}^2/R_{\rm p}$, leading to
\begin{equation}
T_{\rm c} = T_0 \left( {M_{\rm p}\over M_{\rm p0}}\right)^{6\over 3-n} = T_0
\left[{1-z_0\over 1-z}\right]^{6\over 3-n}\;.
\label{tc1}
\end{equation}
This equation describes how temperature of the planet increases as its
metallicity increases.  In the limit $z_0 < z \ll 1$ it can be further
simplified by writing $(1-z_0)/(1-z) \approx 1 + (z-z_0)$, and using the
identity $(1+x)^b \approx \exp(bx)$ valid for $x\ll 1$:
\begin{equation}
T_{\rm c} = T_0  \exp\left[ {6\Delta z\over 3-n}\right]\;,
\label{tc_vs_z3}
\end{equation}
where $\Delta z = z - z_0$. This shows that if $6/(3-n) \gg 1$ then the planet
is very sensitive to addition of grains.

In particular, for di-atomic molecules, $\gamma=7/5$, or $n= 5/2$, which
yields
\begin{equation}
T_{\rm c} = T_0  \exp\left[ {12 \Delta z}\right] = T_0  \exp\left[ {0.18 {\Delta z\over z_\odot}}\right] \;,
\label{tc_vs_z}
\end{equation}
where $z_\odot=0.015$, the Solar metallicity. This suggests that
over-abundance of metals by a factor of $5-10$ in a gas fragment dominated by
molecular hydrogen may significantly increase $T_{\rm c}$, taking the planet
closer to the desired $T_{\rm c}\simgt 2000$~K point at which it can collapse
to much higher densities.

\begin{figure}
\psfig{file=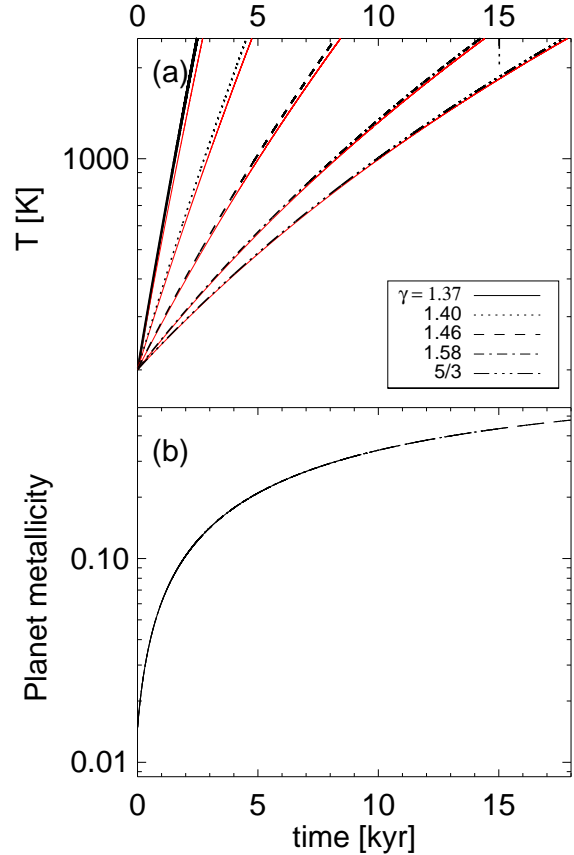,width=0.5\textwidth,angle=0}
  \caption{Evolution of polytropic gas fragments for five different values of
    $\gamma$, as labelled in the figure, when loaded with grains homogeneously
    throughout the cloud. The solid red lines are predictions of the
    analytical model (equation \ref{tc1}), whereas black curves are the
    results of the numerical integration.}
 \label{fig:5gamma}
\end{figure}

Figure \ref{fig:5gamma} show a comparison of the analytical theory (equation
\ref{tc1}) with numerical integrations of the metal loading effect performed
with our code, except that the H/He mix EOS was replaced by the polytropic gas
equation with a fixed $\gamma$. Five values of $\gamma$ are considered, as
labelled on the figure. The metal loading time is $t_z = 300$~years for all of
the curves. The black curves in top panel, fig. \ref{fig:5gamma}a, are
numerical integrations, whereas red lines are equation \ref{tc1}. Figure
\ref{fig:5gamma}b shows the metallicity evolution of the planets, which is
identical, all given by same value of $t_z = 300$ years. The agreement
between the theory and numerical simulations is acceptable to us. The curves
show, just as equation \ref{tc1} does, that the smaller the value of $\gamma$,
the stronger the planet reacts to addition of grains. This is because
$n=1/(\gamma-1)$ edges closer to the unstable value, $n=3$, as $\gamma$
approaches $4/3$ from above, and even a small amount of metals can cause a
significant contraction of the planet.

The same result (equation \ref{tc1}) can be also obtained from a polytropic
sphere relation between planet's radius, mass, and the "constant" $K$, if the
latter is allowed to vary as $z$ changes.  Namely, $K = P/\rho^{1 +1/n} = K_0
[(1-z)/(1-z_0)]^{1+1/n}$ in this case, where $K_0$ is the polytropic constant
at $z=z_0$. Since $K$ is related to entropy per unit mass of the planet, this
provides another interpretation of the metal loading effect: adding metals
decreases entropy of the cloud, just as radiative cooling does, so metal
accretion is effectively a cooling mechanism.

\section{Real H$_2$-dominated polytropes}\label{sec:real}

\begin{figure}
\psfig{file=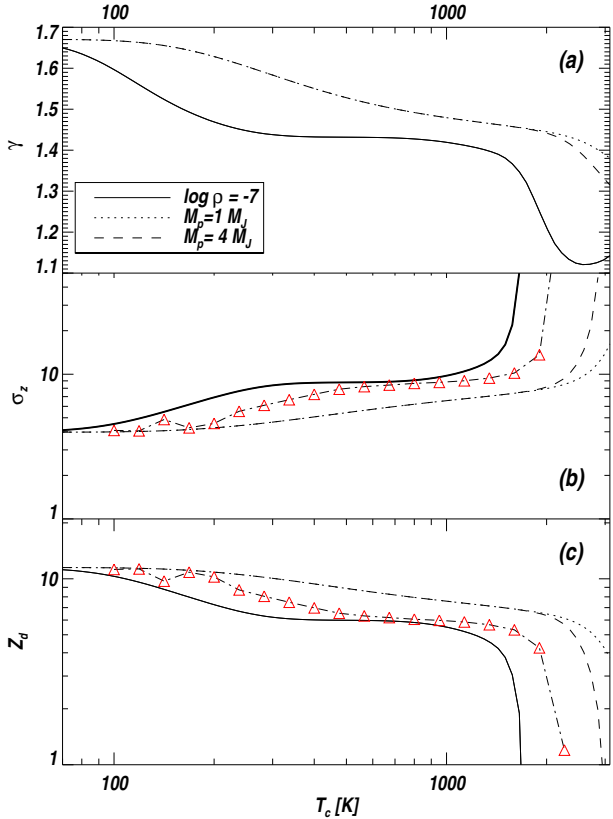,width=0.5\textwidth,angle=0}
  \caption{Panel (a): the ratio of specific heats, $\gamma$, for our Solar
    composition EOS at a fixed gas density, $\rho = 10^{-7}$~g/cm$^3$ (the
    solid curve), and the effective $\gamma$ for two planetary masses as
    defined by equation \ref{egrav}. Middle panel (b): Metallicity exponent
    (equation \ref{sigmaz}) for the same cases as panel (a), plus one directly
    measured from a numerical experiment for $M_{\rm p} = 4\mj$ (dot-dashed
    curve passing through triangles). The larger $\sigma_z$, the more rapidly
    planet contracts as metals are loaded into it. Panel (c): the central
    temperature doubling metallicity $z_{\rm d}$, defined by equation
    \ref{zd1}, for the same cases as in panel (b), and plotted in units of the
    Solar metallicity, $z_\odot$.}
 \label{fig:H2_planets}
\end{figure}

Having explored idealised polytropic models with a fixed value of $\gamma$
(or $n$), we turn to more realistic cases of H/He-dominated pre-collapse
planets. Figure \ref{fig:H2_planets}a shows adiabatic index $\gamma$ versus
gas temperature for a Solar composition H/He plus metals mix for the ideal EOS
used in this paper calculated for a fixed gas density of $\rho = 10^{-7}$
g/cm$^3$. As expected, $\gamma = 5/3$ at $T\simlt 100$~K when rotational and
vibrational degrees of freedom of molecular hydrogen are not yet excited; then
at higher $T$, $\gamma$ drops to $\approx 7/5$ appropriate for diatomic
gas. This persists until $T\sim 1500$~K when $H_2$ dissociation becomes
energetically possible. Due to H$_2$ dissociation at temperatures around
$2000$~K, $\gamma$ plunges to about 1.1.

Clearly, each pre-collapse planet configuration spans a range of densities and
temperatures, from the maximum in the centre to the minimum in the
atmosphere. To relate to the analytical theory derived in \S \ref{sec:theory},
we define effective $\gamma$ and $n$ for a planet by computing the
gravitational potential energy of that planet and then comparing it with that
of a polytropic planet of the same mass and radius, e.g.,
\begin{equation}
E_{\rm grav} = -{3\over 5-n} {G M_{\rm p}^2\over R_{\rm p}}\;,
\label{egrav}
\end{equation}
where $n = 1/(\gamma-1)$. The values of the $\gamma_{\rm eff}$ versus planet's
central temperature are plotted in figure \ref{fig:H2_planets}a for two
planetary masses, $M_{\rm p} = 1 \mj$ and $M_{\rm p} = 4 \mj$. These are very
similar all the way to $T_{\rm c} \approx 2000$~K. As can be expected,
$\gamma_{\rm eff}$ for a planet bears a strong resemblance to the $\gamma(T)$
function (the solid curve), except smoothed out somewhat due to the presence
of a range of temperatures inside a planet, and also shifted to higher
temperatures since $T_{\rm c}$ represents the maximum temperature in a given
planet, whereas $\gamma_{\rm eff}$ is probably related more closely to a mean
temperature in the planet.

Now, the analytical theory developed in \S \ref{sec:theory} predicts how the
central temperature of a polytropic planet varies with its metallicity when
grains are added to the planet, e.g., $T_{\rm c}\propto (1-z)^{-\sigma_{\rm
    z}}$ (equation \ref{tc1}), where we defined "metallicity exponent"
\begin{equation}
\sigma_{\rm z} = - {d \ln T_{\rm c}\over d\ln (1-z)} = {6\over 3-n}\;.
\label{sigmaz}
\end{equation}
The dependence of $\sigma_z$ on central temperature is plotted in figure
\ref{fig:H2_planets}b for the two planetary masses, and also for the solid
curve from figure \ref{fig:H2_planets}a (corresponding to fixed gas density of
$\rho=10^{-7}$ g/cm$^3$).

The dash-dotted curve passing through red triangles in the panel shows $-[d
  \ln T_{\rm c}/d\ln (1-z)]$ actually measured in simulations of planets of
mass $M_{\rm p} = 4 \mj$ loaded with metals. These (short) runs are started
with the planets initialised as polytropic spheres of Solar metal abundance
following the ideal EOS with $T_{\rm c}$ shown on the horizontal axis. The
metals are added to the planets in the same way as described in \S
\ref{sec:intro_runs}, with $t_z = 500$~years (the choice of $t_z$ is
unimportant for this plot). We then measure $\sigma_{\rm z}$ when $z$ varies
by a small increment. To make sure that radiative cooling of the planet, which
also drives evolution of $T_{\rm c}$ in realistic planets, does not corrupt
our measurement of $\sigma_{\rm z}$, we set the opacity multiplier to a very
large number ($10^{10}$) to turn off radiative cooling of the planet for
figure \ref{fig:H2_planets}.

The dash-dotted curve (measured $\sigma_{\rm z}$) should be compared with the
dashed one (predicted). These two are relatively close but do not coincide
exactly, which probably reflects the fact that our definition of $\gamma_{\rm
  eff}$ in equation \ref{egrav} is somewhat arbitrary, and is only one
possible way to define an effective $\gamma$ for a planet. Using the total
energy of the planet to define $\gamma$, for example, gives slightly different
values for $\gamma_{\rm eff}$. Nevertheless, the agreement appears reasonable
and hence the analytical theory has some utility in explaining contraction of
metal-loaded pre-collapse planets.

We also define the central temperature doubling metallicity, $z_{\rm d}$, as
metallicity at which $T_{\rm c}$ is twice its initial value, $T_0$. Given the
definition of $\sigma_{\rm z}$ and equation \ref{tc1},
\begin{equation}
z_{\rm d}  = 1 - \left(1-z_{\rm 0} \right) \, 2^{-1/\sigma_{\rm z}}
\label{zd1}
\end{equation}
where $z_0$ is the initial planet's metallicity. The doubling metallicity in
units of Solar metallicity, and assuming $z_0=z_\odot$, is plotted in figure
\ref{fig:H2_planets}c for the same models as in the two panels above it. We
can see that metal over-abundance of a factor of a few to ten in Solar
metallicity units is required to cause the planet to shrink by a factor of two
in radius (planet's radius is nearly exactly inversely proportional to $T_{\rm
  c}$). The required metal over-abundance is smaller at higher temperatures
because the planet is closer to the unstable value of $\gamma=4/3$ at higher
$T_{\rm c}$.

\section{Grain-dominated planet collapse}\label{sec:grain_dom}

\begin{figure}
\psfig{file=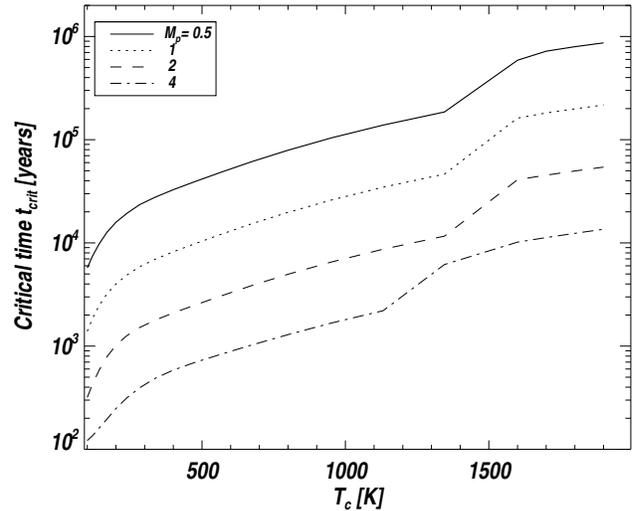,width=0.5\textwidth,angle=0}
  \caption{Critical metal-loading time versus planet's central temperature for
    four different planetary masses and interstellar grain opacities, as
    labelled in the box in the figure. Metals must be added to the planet
    rapidly, so that $t_z < t_{\rm crit}$, in order for the metal loading to
    dominate over radiative cooling as a contraction mechanism.}
 \label{fig:tz_crit}
\end{figure}

Having studied the contraction rates of planets loaded by metals, we can now
ask the question of when such a contraction is faster than that due to
radiative cooling. This happens when the rate of central temperature increase
due to radiative cooling rate is equal to the that due to metal loading. The
latter can be found via
\begin{equation}
{d T_{\rm c} \over dt} = T_{\rm c} {d \ln T_{\rm c}\over d\ln (1-z)}\;{d\ln (1-z) \over dt}\;.
\end{equation}
Now, using equation \ref{sigmaz} and definition $1-z = M_{\rm gas}/(M_{\rm
  gas} + M_{\rm z})$, where $M_{\rm gas}$ and $M_{\rm z}$ and the total mass
of gas and metals in the planet, we write
\begin{equation}
{d\ln (1-z) \over dt} = - {1\over M_{\rm p}} {d M_{\rm z} \over dt} = -
{M_{\rm z}(0)\over M_{\rm p} t_{\rm z}}
\label{dzdt1}
\end{equation}
The condition of the two contraction rates being equal to each other reads
\begin{equation}
{T_{\rm c} \over t_{\rm rad}} = T_{\rm c} \sigma_{\rm z} \;
{z_0\over t_{\rm z}}
\label{eq_rates}
\end{equation}
where we set $z=z_0$. This defines the critical metal loading time scale,
$t_{\rm crit}$, such that if metals are supplied to the planet on a time scale
shorter than $t_{\rm crit}$ then the planet contracts mainly due to the
increasing metal content rather than radiative cooling. According to equation
\ref{eq_rates}, the critical time is
\begin{equation}
t_{\rm crit} = t_{\rm rad} \sigma_{\rm z} z_0\;,
\label{tcrit1}
\end{equation}
Figure \ref{fig:tz_crit} plots the critical metal loading time versus planet's
central temperature for the same range of planetary masses as in figure
\ref{fig:tcool}. The curves showing $t_{\rm crit}$ are qualitatively similar
to those showing $t_{\rm rad}$ for obvious reasons save for the absolute
values, and the factor $\sigma_{\rm z}$ that is a function of $T_{\rm c}$.

This figure shows that loading the planet by metals at a typical $t_{\rm z}$
of a few thousand years, as estimated in \S \ref{sec:mdot_grain}, is not
likely to affect relatively massive planets, $M_{\rm p}\simgt 2 \mj$, at the
beginning of their evolution, but may affect them when they contract towards
$T\simgt 1000$~K. This finding is consistent with the right panels of figure
\ref{fig:exp1}, in which the central temperature evolution of metal-loaded
planets fed by metals at different rates was shown. The figure showed that
early evolution of planets, when $T_{\rm c}\simlt 500$~K, is
indistinguishable for different $t_{\rm z}$, at least in a rather broad range
of $t_{\rm z}$, from 250 years all the way to $t_{\rm z}=\infty$. This can now
be understood in terms of the radiation-dominated phase of the planet's
contraction: early on the planet contracts radiatively more rapidly than it
does due to metal loading, even at the shortest $t_{\rm z}$ explored in the
figure. At later times, however, depending on $t_{\rm z}$, the planets go off
the fixed metallicity (solid curve in figure \ref{fig:exp1}c) track and
accelerate onto the metal-dominated part of their evolution.

On the other hand, figure \ref{fig:tz_crit} predicts that evolution of lower
mass gas fragments may be much more sensitive to deposition of metals. For
example, for $M_{\rm p} =0.5\mj$, even $t_{\rm z} = 10^4$ years is short
enough to dominate fragment's contraction. One preliminary conclusion from
this, which clearly needs to be explored more rigorously in a realistic
evolving protoplanetary disc setting, is that lower mass giant planets formed
by GI/TD may be more metal rich. Physically, it takes far longer for these
planets to contract (because their radiative cooling times are much longer,
see figure \ref{fig:tcool}). For this reason, they are likely to have a lower
central temperature for longer than their massive cousins. Based on the theory
developed above and figure \ref{fig:H2_planets}, their contraction due to
metal loading is therefore slower, and hence they may be able to absorb more
metals before they collapse.

\begin{figure*}
\begin{minipage}{3.2in}
\psfig{file=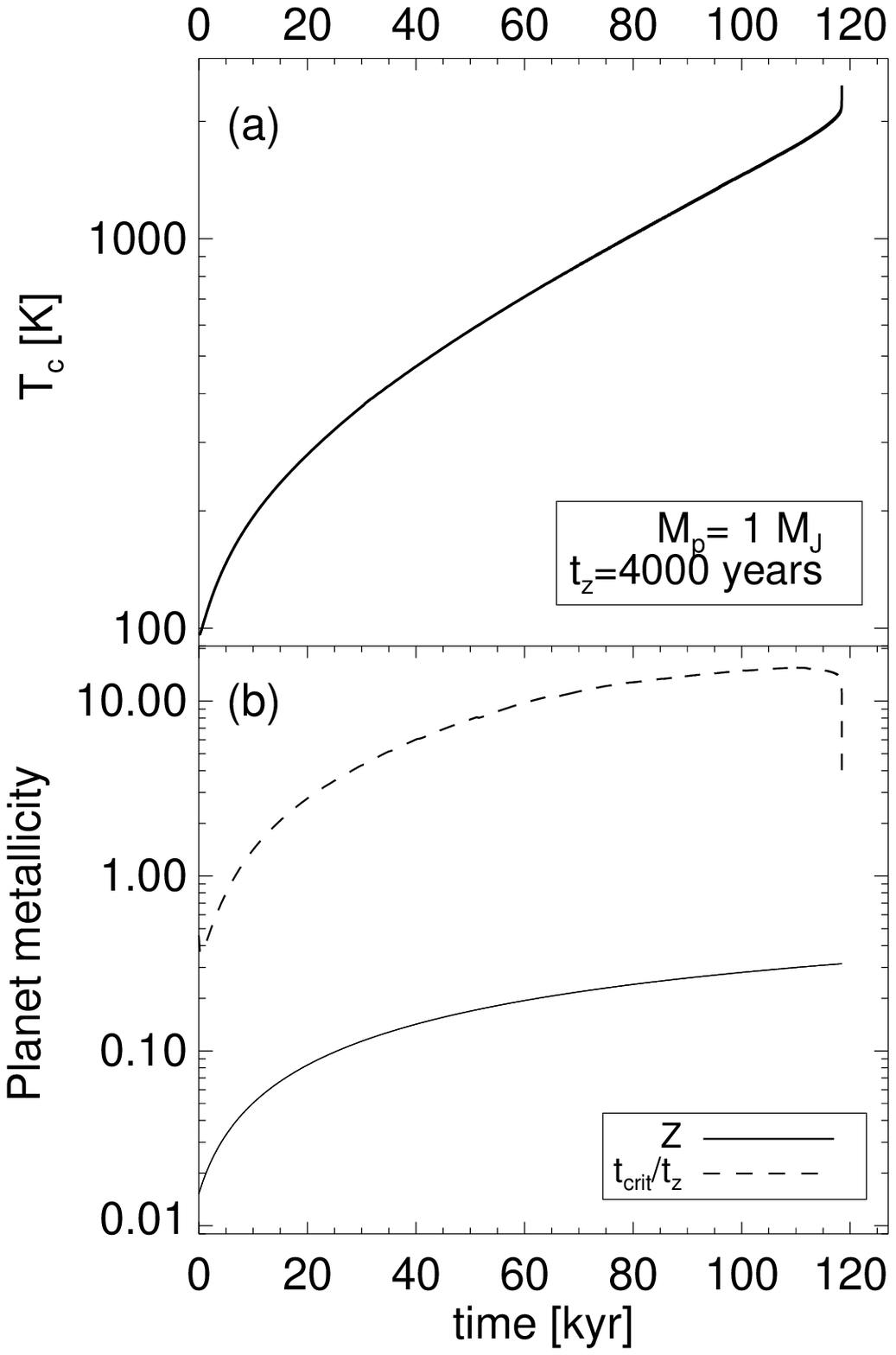,width=0.99\textwidth,angle=0}
\end{minipage}
\begin{minipage}{3.2in}
\psfig{file=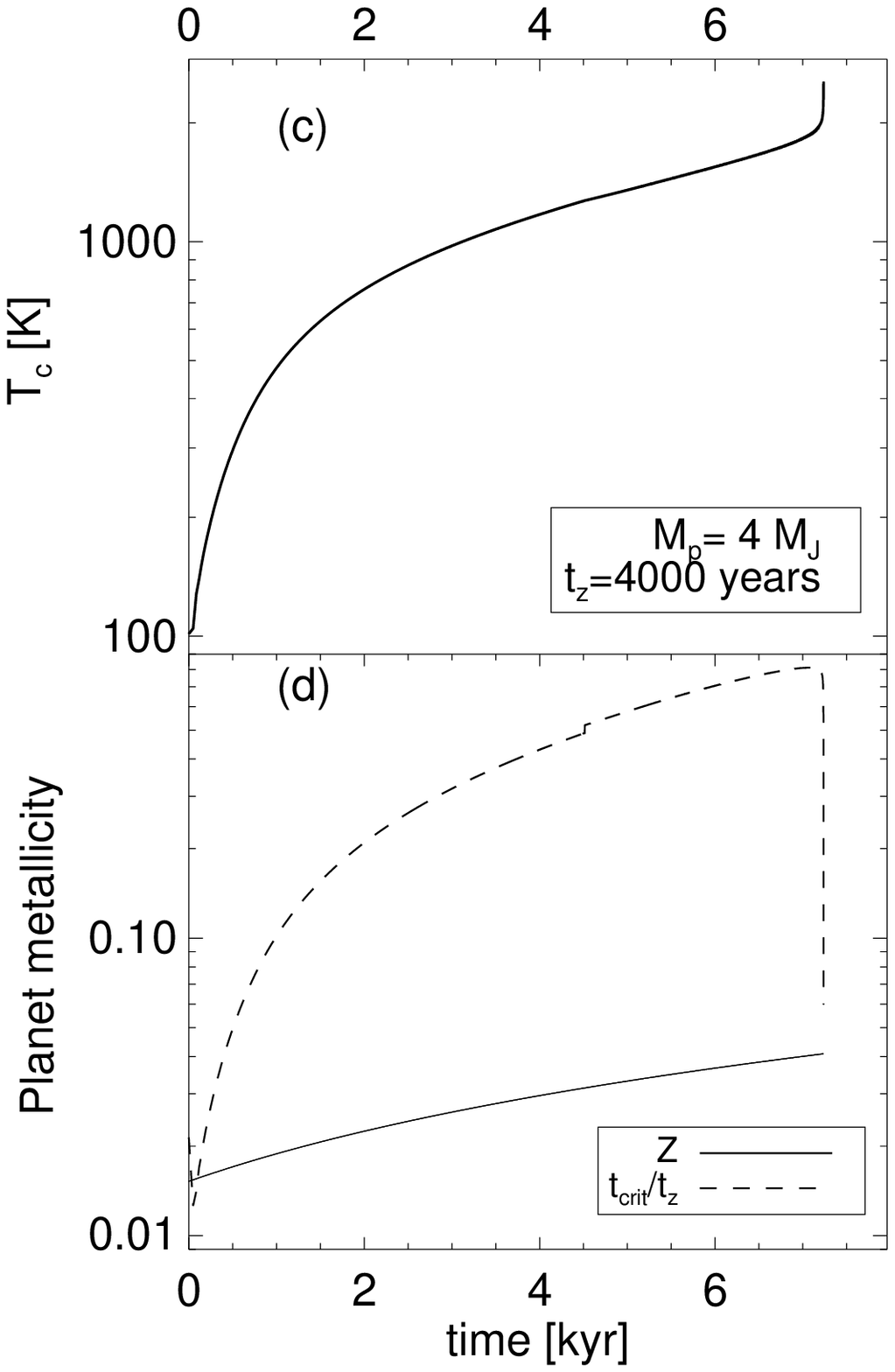,width=0.99\textwidth,angle=0}
\end{minipage}
  \caption{The central temperature (top panels, a and c) and metallicity
    (bottom panels, b and d) versus time for planets with mass $M_{\rm p} = 1
    \mj$ and $M_{\rm p} = 4 \mj$ (left and right panels, respectively). The
    dashed curve in the bottom panels show the ratio of the critical metal
    loading time scale to the actual value of $t_{\rm z}$ ($4000$~years for
    both runs). The dashed curves show that the evolution of the $M_{\rm p} =
    1 \mj$ is dominated by metal loading, whereas that of the $M_{\rm p} =
    4\mj$ is dominated by radiative cooling.}
 \label{fig:M1M4}
\end{figure*}

While a full scale investigation of this prediction is outside the scope of
our paper, figure \ref{fig:M1M4} compares the evolution of two planets,
$M_{\rm p} = 4\mj$, which was already shown in figures \ref{fig:exp1}c,d with
the long dash curves, while the other planet has a lower mass, $M_{\rm p} =1
\mj$. The solid lines in figures \ref{fig:M1M4}a,b show the evolution of the
central temperature and metallicity, $z$, respectively, for $M_{\rm p} =1
\mj$. The dashed curve in figure \ref{fig:M1M4}b show the ratio of the
critical metal loading time scale to the actual value of $t_{\rm z}$
($4000$~years for both runs). Except for the earliest times, that ratio is
always greater than unity, suggesting that evolution of this planet is nearly
always metal dominated. This planet collapses at $t=0.12$~Myr almost entirely
due to metal loading rather than radiation. By the time the planet collapses,
its metallicity is very high, $z\approx 0.3$, e.g., some 20 times higher than
$z_\odot$.

The evolution of the more massive planet is distinctly different. Since such
planets are cooling much more rapidly (see fig. 1), metal loading is a
relatively minor effect in this case, with $t_{\rm crit}/t_{\rm z} < 1$
everywhere (see figure \ref{fig:M1M4}d). The evolution of a massive planet is
hence radiation-dominated. The critical loading time $t_{\rm crit}$ is
  calculated using equation \ref{tcrit1}, where the radiative time is found at
  each time step using the definition from equation \ref{trad}, the
  metallicity exponent $\sigma_z$ is computed using equation \ref{sigmaz} and
  definition of $n$ given by equation \ref{egrav}.

\section{Discussion}\label{sec:discussion}

In this paper we have shown that accretion of pebbles from the surrounding
protoplanetary disc is a surprisingly efficient way for pre-collapse H$_2$
dominated planets to contract and eventually collapse in high opacity
regime. The simplest way to understand this result is to (1) realise that a
gentle sedimentation of grains onto a planet brings in additional
gravitational potential energy but not thermal or kinetic energy, so that the
total energy of the cloud becomes more negative. (2) The reason for a
surprising sensitivity of molecular pre-collapse planets to metals is the fact
that the total energy of a polytropic cloud with index $n$ (equation
\ref{etot1}) with $n\rightarrow 3$ is very small compared to $G M_{\rm
  p}^2/R_{\rm p}$ (e.g., $(3-n)/(5-n) = 0.2$ for $n=5/2$), so it takes only a
little extra weight in metals to topple the planet over into the unstable
H$_2$ dissociation regime.

The implications of our results for TD hypothesis and GI-born planets are
potentially significant. First of all, metal loading makes it much more likely
that these planets are able to contract and collapse despite a rapid inward
migration. This may help to resolve the challenges to forming GI planets at
short separations by migration of fragments born at separations $\sim 100$~AU
emphasised by \cite{ZhuEtal12a} and \cite{VazanHelled12}. Secondly, metal
loading process may explain why giant planets are over-abundant in metals
\citep[e.g.,][]{MillerFortney11}. Finally, TD/GI planet survival may be
enhanced at high metallicities since the rates of pebble accretion are higher,
making the planets collapse sooner, while planet migration rates are unlikely
to depend on disc metallicity strongly. This may produce a positive giant
planet frequency -- host star metallicity correlation as needed to explain
observations \citep{FischerValenti05}. These issues will be addressed in a
separate forthcoming publication where the disc-planet interaction is modelled
in detail.

As discussed in the Introduction, low opacity in pre-collapse GI clumps
  may be another way to speed up the collapse \citep{HB11}, but it is not
  clear if opacity in the clumps is indeed much smaller than the interstellar
  one. \cite{DD05} argued that grain aggregate fragmentation in higher speed
  collisions is required by observations of T Tauri stars to maintain a high
  abundance of small grains despite the simultaneous presence of grains as
  large as $\sim1 $~cm in size. Furthermore, the low dust opacity picture
  would predict a very efficient formation of giant planets at low
  metallicities, which contradicts observations \citep{FischerValenti05}.  

We also found that the response of giant planets to metal loading is a strong
function of their mass. More massive planets are less likely to be influenced
by metal accretion since they cool much more rapidly than their lower mass
cousins (see fig. \ref{fig:M1M4}). This gives us hope that there may be strong
trends in predictions of the TD model with fragment's mass that could
hopefully be used to contrast it with observations of exoplanets.

The main shortcomings of our numerical approach here are (i) the
simplification in the EOS that neglects metals' contribution to the internal
energy and pressure of the mix (see Appendix), and (ii) assuming a homogenous
spreading of the grains inside the planet. Both of these are made in order to
keep our treatment as transparent as possible and to enable a comparison to
the analytical model developed in \S \ref{sec:theory}. Furthermore,
preliminary results of simulations with a more complete EOS show that this
only strengthens the tendency of the fragments to collapse due to metal
loading because grain vaporisation must occur before the fragment collapses
(as $T\simgt 2000$~K at that point), and this process takes extra energy from
the gas due to the latent heat of vaporisation \citep[e.g.,][]{PodolakEtal88},
cooling the fragment yet more.  The assumption that grains are well mixed with
gas is reasonable if convection stirs the grains up quickly
\citep[e.g.,][]{HB11}, mixing newly accreted grains throughout the planet
efficiently. Nevertheless, we shall endeavour to relax these simplifications in
future publications.

\section{Acknowledgments}

Theoretical astrophysics research at the University of Leicester is supported
by a STFC Rolling grant. The author thanks Richard Alexander for useful
discussions.  This work used the DiRAC Complexity system, operated by the
University of Leicester, which forms part of the STFC DiRAC HPC Facility
(www.dirac.ac.uk). This equipment is funded by a BIS National E-Infrastructure
capital grant ST/K000373/1 and DiRAC Operations grant ST/K0003259/1. DiRAC is
part of the UK National E-Infrastructure.

\bibliographystyle{mnras}

\appendix

\section{Equations}\label{appendix1}

The equations are solved in spherical symmetry, with {\em gas} mass $M_{\rm
  g}$, the primary coordinate counted from the centre of the planet, used to
set a staggered grid. The numerical time integration procedure is based on the
Lagrangian scheme ``lh1'' presented in \S 6.2 of \cite{BodenheimerBook}. In
the one-fluid limit the equations are
\begin{eqnarray}
{1\over \rho} = 4\pi r^2 {d r \over d M}
\label{rho_def}\\
\frac{d v}{dt} = - 4\pi r^2 \frac{\partial P}{\partial M} - \frac{GM(r)}{r^2} \;,\\
\label{rad_f}
\frac{d u}{dt} = - 4\pi P \frac{\partial (r^2 v)}{\partial M} - 4\pi
\frac{\partial \left(r^2 F(r)\right)}{\partial M}\;
\label{main_eqs}
\end{eqnarray}
where $v = dr/dt$ is gas velocity, $M(r) = M_{\rm g}(r) + M_{\rm z}(r)$ is the
total (gas + grains) mass enclosed within radius $r$, $\rho$ and $P$ is the
total gas density and pressure at radius $r$. Note that since $1-z$ is the
mass fraction of gas (H/He in the context of this paper), the total mass can
be written as
\begin{equation}
M(r) = {M_{\rm g}(r)\over 1-z}\;,
\label{mdef}
\end{equation}
and similarly, total density $\rho = \rho_{\rm g}/(1-z)$ at all radii inside
the planet. $F$ is the radial energy flux, and is equal to either the
radiative flux, $F_{\rm rad}$, given by the classical radiation flux,
\begin{equation}
F_{\rm rad} = - \frac{4 a_{\rm rad} cT^3 }{3\kappa(T)}\,4\pi r^2\,
\frac{\partial T}{\partial M}\;,
\label{flux}
\end{equation}
where $a_{\rm rad}$ is the radiation constant, or $F_{\rm cov}$, the
convective energy flux if the conditions for convective instability are
satisfied. The latter is modelled according to the mixing length theory. The
numerical scheme is explicit, with the time step limited by the Courant
condition. An artificial viscosity is used to capture shocks.  The equation
of state (EOS) that we use provides gas pressure and internal energy as a
function of total density $\rho_{\rm tot}$, metallicity $z$ and gas
temperature $T$.  Since gas temperature $T$ is not explicitly present in
equations \ref{rho_def} to \ref{main_eqs}, at the end of each time step, which
yields the values of $P$, $\rho$, $u$ and $z$, at every mass zone, an
iterative procedure is implemented to find the corresponding $T$.

The EOS of H/He mix with astrophysical metals is in general very complicated,
and one needs an extensive chemical network to follow phase transitions and
interactions between different chemical species. For example,
\cite{HoriIkoma11} considered a number of compounds formed by elements C and O
with hydrogen, and found that thermodynamical properties of gas change
strongly at high metallicity ($z\rightarrow 1$). \cite{VazanEtal13} used
SiO$_2$ to model "rock" in their EOS.  Lacking such proprietary EOS tables, we
use a simpler approximation in which grains/metals do {\em not} contribute to
the gas pressure or the internal energy of the gas other than by increasing
the mass of the mixture. This approximation is excellent while metals are all
in the grain phase because the energy in Brownian motion of the grains is
negligible compared with that of the much lighter H/He molecules or atoms. The
approximation is less well justified when grains are vaporised and the
constituent molecules are broken down at higher temperatures. However,
including the latent heat of grain vaporisation and metal molecules
dissociation would only strengthen conclusions of our paper since the extra
heat required for these processes would have been taken from the thermal
energy of the fragment and would thus lead to its faster collapse. The
simplified EOS we use therefore presents the minimum metal loading effect, and
the more complicated forms of it must make it even stronger.

The H/He part of the EOS are modelled as that of an ideal H/He gas mixture
which includes hydrogen molecules rotational and vibrational degrees of
freedom \citep[calculated as in][]{BoleyEtal07}, and H$_2$ dissociation
\citep[cf. more technical detail in][]{Nayakshin11b}, as well as ionisation of
hydrogen atoms. Ionisation of He atoms is not important for temperatures of
interest, and is hence neglected.

The EOS of the toy model of an ideal polytropic gas planets that we use in \S
\ref{sec:theory} is $P = K \rho_{\rm tot}^{\gamma}$, where $\gamma$ is a fixed
adiabatic index, and $K$ is a constant related to the entropy of the gas. The
internal energy is then given by
\begin{equation}
u = {P\over \rho}\,{1\over \gamma-1}\;.
\label{up0}
\end{equation}
The mean molecular weight of the mixture is calculated as
\begin{equation}
{1\over \mu} = {1-z\over \mu_{\rm g}} + {z\over \mu_{\rm dust} }= {1-z\over \mu_{\rm g}}\;,
\label{mua0}
\end{equation}
where $\mu_{\rm g}$ is the mean molecular weight of the H/He mixture (which
depends on the fraction of H atoms in H$_2$ molecules, atomic H and ionised H;
He atoms are assumed to be not ionised). The last step in equation \ref{mua0}
is possible because the mean molecular weight of dust particles is orders of
magnitude larger than $\mu_{\rm g}$.

\end{document}